\documentclass[prb,twocolumn]{revtex4-1}

\usepackage{mathtools}
\usepackage{amsmath,bm}
\usepackage{amssymb}
\usepackage{subfigure}
\usepackage{setspace}
\usepackage{dcolumn}
\usepackage{color}

\usepackage[hidelinks]{hyperref}

\begin{document}

\title{Experimental observation of temporal pumping in electro-mechanical waveguides}

\author{Yiwei Xia$^{a}$, Emanuele Riva$^{b}$, Matheus I. N. Rosa$^{c}$, Gabriele Cazzulani$^{b}$, Alper Erturk$^{a}$, Francesco Braghin$^{b}$ and  Massimo Ruzzene$^{c}$}
\affiliation{ $^a$ School of Mechanical Engineering, Georgia Institute of Technology, Atlanta GA 30332}
\affiliation{ $^b$ Department of Mechanical Engineering, Politecnico di Milano, Italy}
\affiliation{ $^c$ Department of Mechanical Engineering, University of Colorado Boulder, Boulder CO 80309}
%\thanks{}

\date{\today}

\begin{abstract}
We experimentally demonstrate temporal pumping of elastic waves in an electromechanical waveguide. An aluminum beam covered by an array of piezoelectric patches connected to shunt circuits with controllable resistances enables the spatial and temporal control of the beam's stiffness. The spatial modulation produces non-trivial edge states, while a smooth temporal variation of the modulation phase drives the transfer of edge states from one boundary of the waveguide to the other in a controllable manner. This characteristic behavior for a topological pump is here demonstrated for the first time in a continuous elastic waveguide. The framework presented herein opens new avenues for the manipulation and transport of information through elastic waves, with potential technological applications for digital delay lines and digitally controlled waveguides. Our study also explores higher dimensional topological physics using time as a synthetic dimension in electromechanical systems.
\end{abstract}

\maketitle

%\section{Introduction}\label{Introduction}

The transport of information along one-dimensional (1D) waveguides is key to numerous technological applications, but is generally limited by two main factors: $(i)$ fixed propagation speeds and associated wave dispersion that are determined by the properties of the medium, and $(ii)$ scattering and localization of the propagating signal at defects and imperfections. The study of topological insulators has opened new pathways to overcome these limitations, sparkling broad interest across different realms of physics, including quantum,~\cite{hasan2010colloquium} electromagnetic,~\cite{lu2014topological,khanikaev2013photonic} acoustic~\cite{PhysRevLett.114.114301,fleury2016floquet,lu2017observation} and elastic~\cite{mousavi2015topologically} media. Robust waveguiding along edges and interfaces of two-dimensional (2D) domains have been demonstrated in different elastic and acoustic platforms, exploiting analogies with the \textit{Quantum Hall Effect} (QHE),~\cite{klitzing1980new,thouless1982quantized,prodan2009topological,wang2015topological,nash2015topological,souslov2017topological,mitchell2018amorphous,chen2019mechanical} the \textit{Quantum Spin Hall Effect} (QSHE)~\cite{mousavi2015topologically,susstrunk2015observation,pal2016helical,PhysRevB.98.094302,chaunsali2018subwavelength,PhysRevX.8.031074} and the \textit{Quantum Valley Hall Effect} (QVHE).~\cite{pal2017edge,vila2017observation,liu2018tunable,liu2019experimental} These works and the references therein illustrate a variety of strategies for robust wave transport, which generally require 2D domains, and occur at fixed (non-tunable) speeds.

A recent line of work exploits synthetic dimensions to explore higher dimensional topological effects in lower dimensional systems.~\cite{qi2008topological,kraus2016quasiperiodicity,ozawa2016synthetic,lee2018electromagnetic} For example, edge states commonly attributed to (2D) QHE systems have been illustrated in 1D periodic~\cite{alvarez2019edge,rosa2019edge} and quasiperiodic~\cite{kraus2012topological,apigo2019observation,ni2019observation,Pal_2019,xia2020topological} systems, while 4D and 6D Quantum hall phases have been observed in 2D~\cite{zilberberg2018photonic,lohse2018exploring} and 3D~\cite{petrides2018six,lee2018electromagnetic} lattices. In this context, topological pumping emerges as a phenomenon of particular interest, whereby transitions of edge states from one boundary to another of a 1D system are induced by parametric variations along one additional (synthetic) dimension, either spatial~ \cite{kraus2012topological,verbin2015topological,zilberberg2018photonic,lohse2018exploring,rosa2019edge,riva2020edge} or temporal.~\cite{nakajima2016topological,lohse2016thouless,grinberg2020robust,chen2019mechanical,brouzos2019non,longhi2019topological,riva2020adiabatic} A temporal pump embodies a 2D topological effect that governs the robust energy transport in systems of a single spatial dimension. While the concept is very promising and supported mostly by theoretical investigations,~\cite{nassar2018quantization,chen2019mechanical,brouzos2019non,riva2020adiabatic} its experimental realization for elastic waves has so far been elusive. Notable recent studies include the temporal pumping illlustrated in a dimerized magneto-mechanical system emulating the Su-Schreefer-Heeger (SSH) model,~\cite{grinberg2020robust} and the mapping of egde state transitions in reconfigurable quasiperiodic acoustic lattices.~\cite{ni2019observation} These contributions highlight the potential of temporal pumping for robust wave transport, but suggest that further efforts may be required towards implementations in compact and modular configurations, which can be potentially scaled down for on-chip applications.

Towards bridging this gap, we leverage prior work on topological pumping through spatial stiffness modulations,~\cite{rosa2019edge,riva2020edge} on temporal modulation for frequency-selective filtering, and on spatio-temporal modulations for non-reciprocal wave motion.~\cite{trainiti2019time,marconi2020experimental} We implement a digitally controllable temporal pump for elastic waves propagating in a continuous waveguide, whereby adiabatic modulation of the bending stiffness is obtained through the control of the electrical impedance of circuits shunting piezoelectric arrays.~\cite{marconi2020experimental} Spatial modulation produces edge states,~\cite{rosa2019edge,riva2020edge} while the smooth temporal variation of the modulation phase leads to the transition of the edge states from one boundary of the waveguide to the other, which is the hallmark of topological pumping. We provide the first experimental demonstration of adiabatic temporal pumping in a continuous waveguide through an experimental set-up that is suitable for miniaturization and for device-level implementation. The results proposed herein open avenues for wave transport through topological phenomena that employ time as a synthetic dimension, and suggest how adiabatic pumping may enable the transfer of a signal along a waveguide with tunable arrival times and delays, which is relevant to the design of controllable acoustic delay lines.

%\section{Analysis of edge states in elastic beams with shunted piezoelectric array}\label{theorysec}

\begin{figure}[t!]
\includegraphics[height=0.25\textwidth]{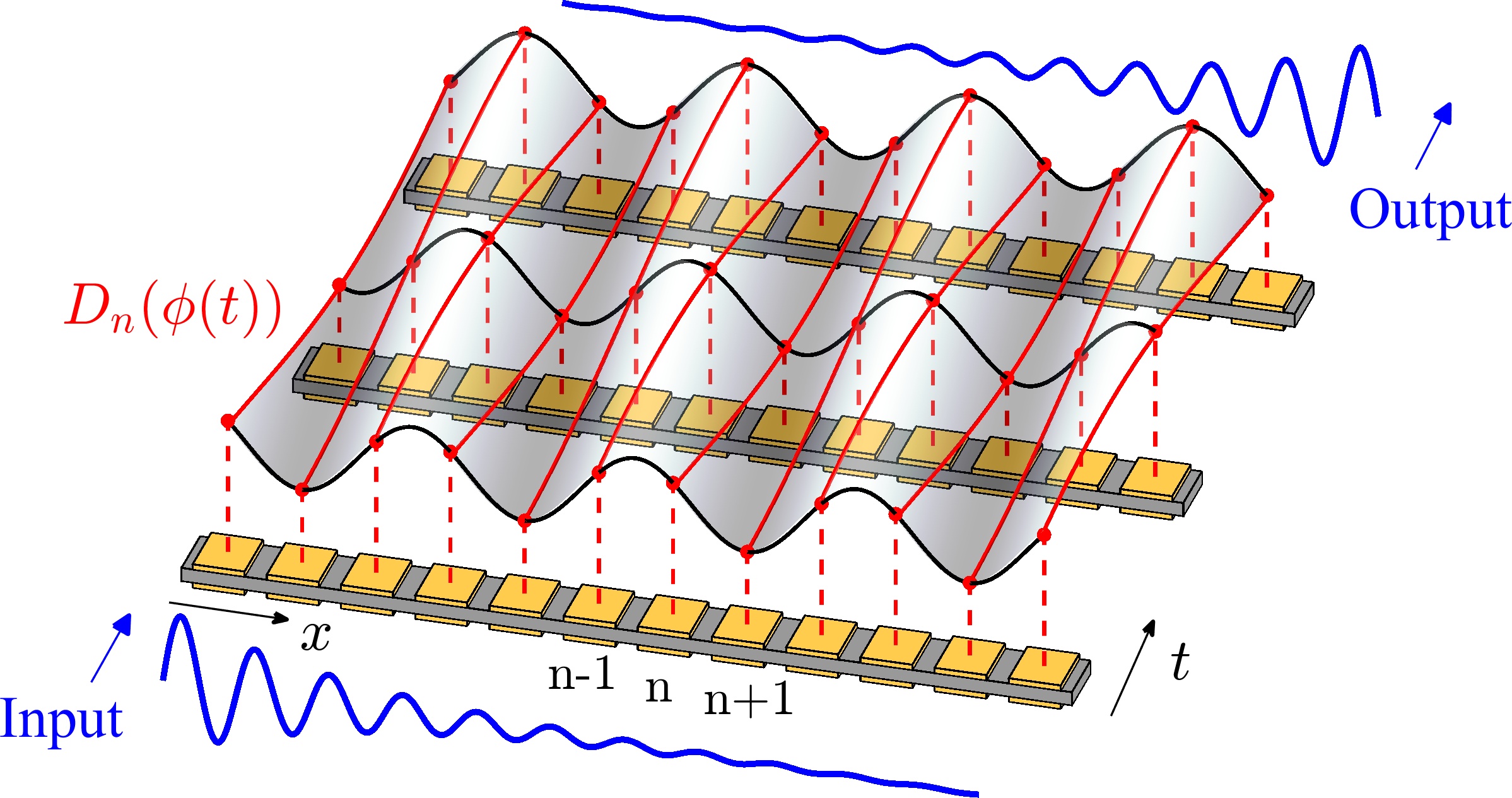}
\centering
\caption{Concept of temporal pumping implemented in electromechanical elastic beam. The equivalent stiffness $D_{n}(\phi)$ at the location of the n-th piezoelectric patches (red lines) is obtained by sampling the surface $D(x,\phi)=D_{0}[1+\alpha\cos(2\pi\theta x+\phi)]$ at $x_n=n$ (Ref.~\cite{rosa2019edge}). The spatial stiffness modulation with slowly varying temporal phase $\phi(t)$ induces the transition of the left-localized edge state (input) into a right-localized state (output).}
\label{Fig1}
\end{figure}

We consider an elastic beam (gray solid in Fig.~\ref{Fig1}) where spatio-temporal modulation is induced by an array of piezoelectric patches (yellow), shunted through negative capacitance (NC) circuits.~\cite{marconi2020experimental} The NC shunts modify the equivalent bending stiffness $D$ according to the following modulation:
\begin{equation}\label{modEQ}
D_{n}(\phi) = D_{0}[1+\alpha\cos(2\pi n \theta+\phi)],
\end{equation} 
where $D_n$ is the contribution to the bending stiffness at the location of the $n^{\text{th}}$ patch (Fig.~\ref{Fig1}). This spatial stiffness modulation produces edge states localized at one of the boundaries of the waveguide depending on the assigned value of the modulation phase $\phi$.~\cite{rosa2019edge} An adiabatic temporal modulation of the phase $\phi(t)$ drives a left-localized edge state (input) across the waveguide producing a right-localized state (output), thus implementing topological pumping. 

Wave motion along the waveguide is predicted by employing Euler-Bernoulli beam theory,~\cite{graff2012wave} which describes the transverse harmonic motion at $w(x,\omega)$ of the waveguide through the following governing equation:
\begin{equation}\label{govEQ}
[D(x)w_{,xx}]_{,xx} = \omega^2 m(x) w(x,\omega)
\end{equation}
where $()_{,x}$ denotes a derivative with respect to $x$, while $m$ is the linear mass the beam. Because of the presence of the patches, stiffness and inertia properties are periodic functions of $x$, and can be expressed as:
\begin{eqnarray}\label{eq:periodic properties}
D(x)=D_b+\sum_n D_n H(x-n x_p,l_p), \\
m(x)=m_b+\sum_n m_p H(x-n x_p,l_p)
\end{eqnarray}
where $D_n$ is given in eq.n~\eqref{modEQ}, $D_b,m_b$ respectively denote the bending stiffness and linear mass of the base beam, and $m_p$ is the increase in linear mass at the locations of the patches. Aslo in eqn.~\eqref{eq:periodic properties}, $H(\cdot)$ is a unit step function centered at location $n x_p$ and of length $l_p$.

We consider a modulation with $\theta=1/3$ in Eqn.~\eqref{modEQ}, resulting in a periodic beam, whose period $L_c=72$ mm comprises 3 piezoelectric elements of length $l_p=22$ mm, that are $2$ mm apart. The variable resistance NC shunts produce a stiffness modulation that is quantified by a value of $\alpha=0.172$. The estimation of these values is based on the procedures described in Supplemental Materials (SM),~\cite{SM} where details about the system geometrical and physical parameters are found.

\begin{figure*}[ht!]
	\centering
	\subfigure[]{\includegraphics[height=0.275\textwidth]{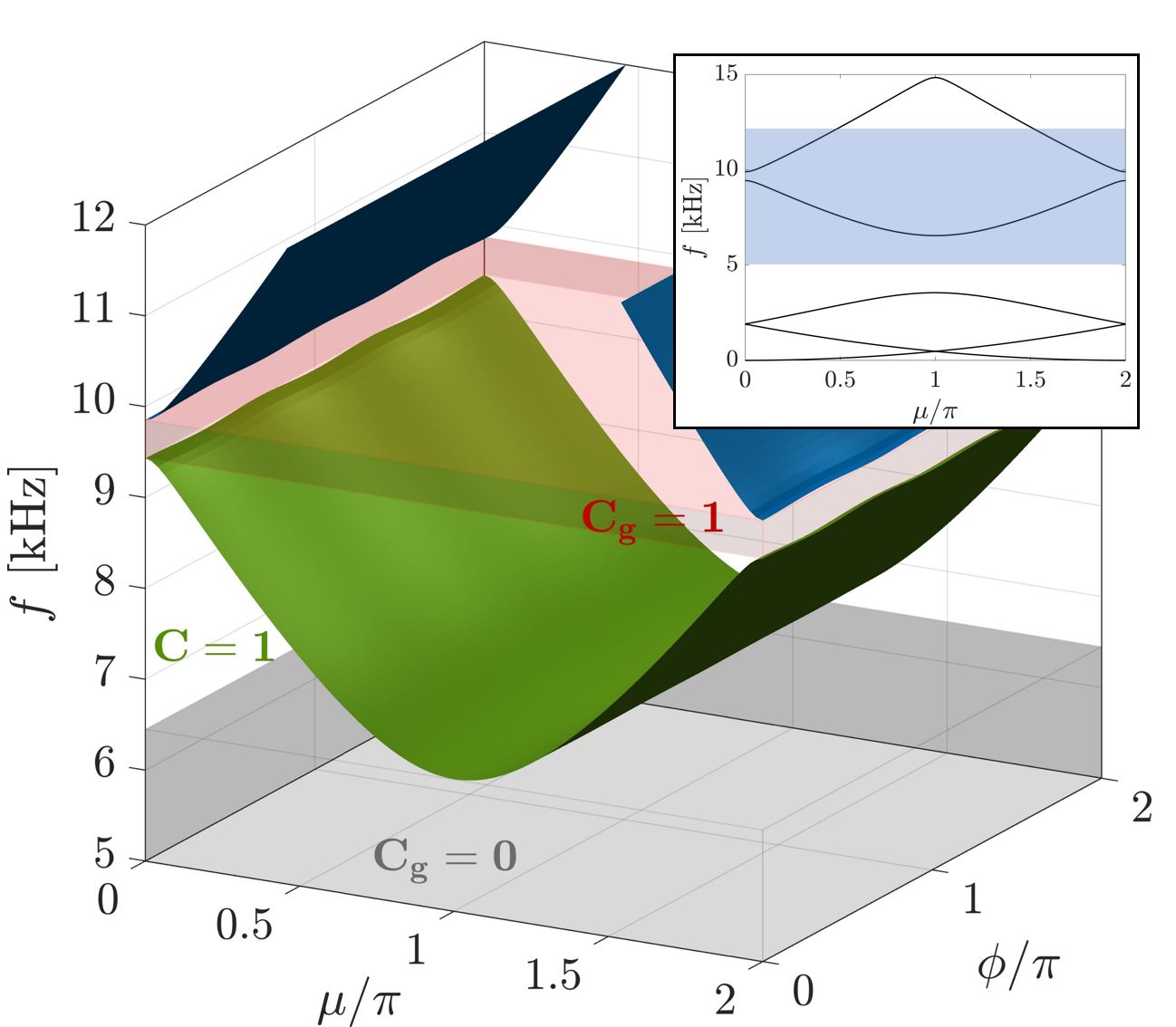}\label{Fig2a}}
	\subfigure[]{\includegraphics[height=0.275\textwidth]{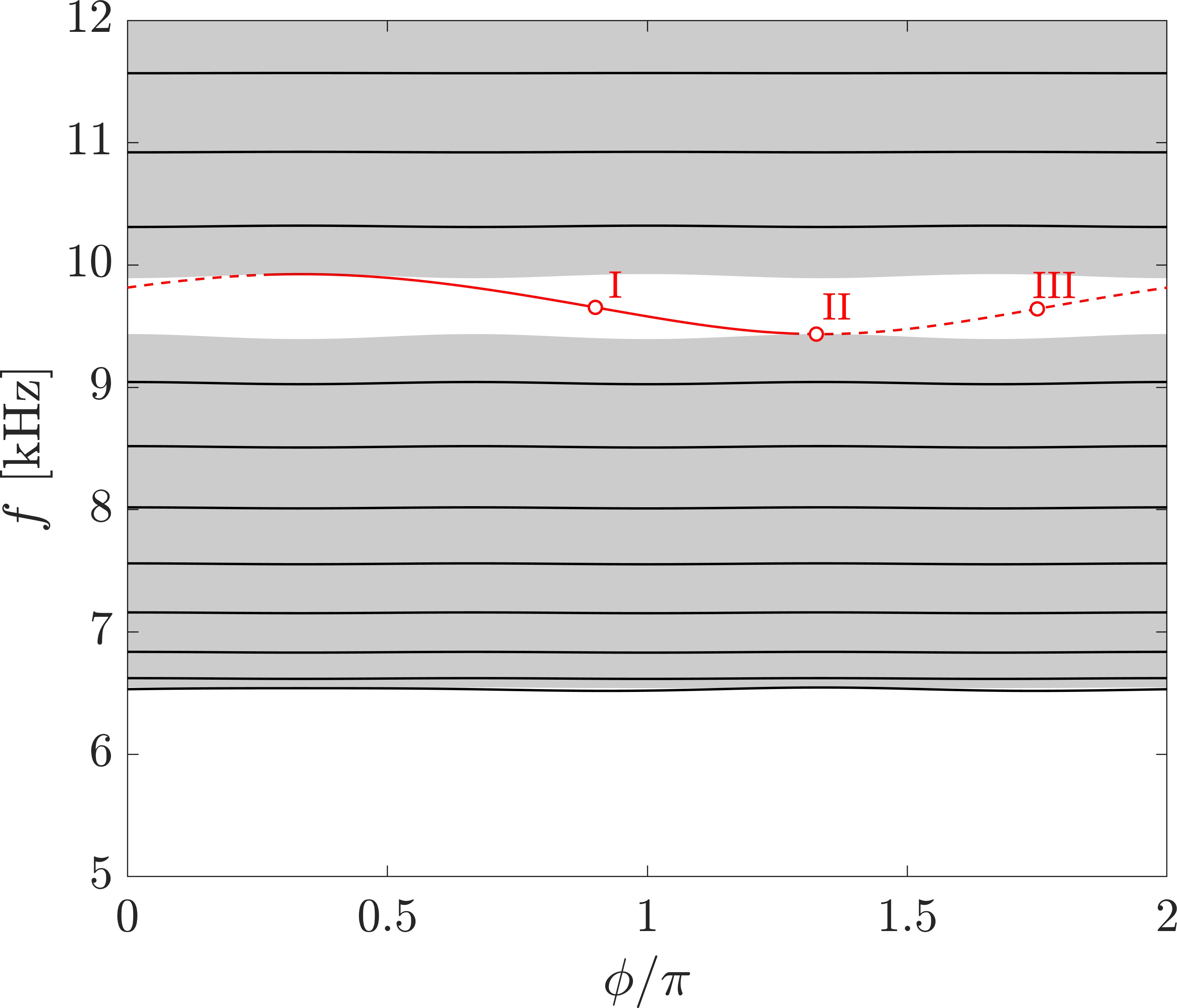}\label{Fig2b}}
	\subfigure[]{\includegraphics[height=0.275\textwidth]{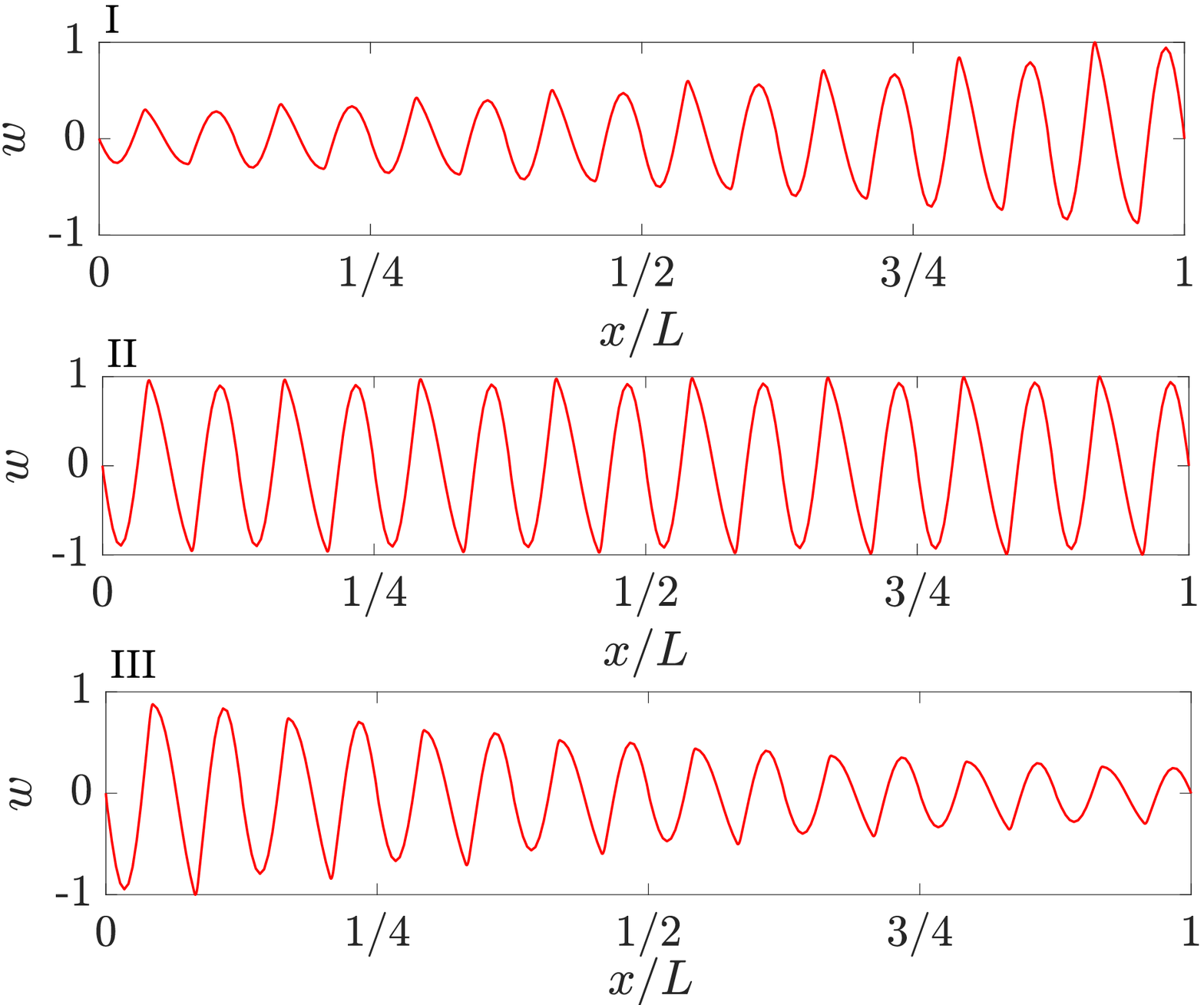}\label{Fig2c}}
	\caption{Dispersion properties and edge states for a beam with equivalent stiffness modulation $D_{n}(\phi) = D_{0}[1+\alpha\cos(2\pi\theta n+\phi)]$. (a) Dispersion surfaces as a function of $\mu$ and $\phi$ with information on Chern numbers and gap labels. The inset displays the first five bands for $\phi=0$, highlighting the frequency range considered for the surface plot (shaded blue region). (b) Eigenfrequencies for a finite beam as a function of $\phi$ (black lines) superimposed to the bulk bands (shaded gray regions), where an edge state (red lines) spans the non-trivial gap with $C_g=1$. (c) States corresponding to the points marked in (b) showing examples of right-localized mode (I), bulk mode (II) and left-localized mode (III).}
	\label{Fig2}
\end{figure*}

We first investigate the dispersion properties $\omega(\kappa,\phi)$ of the modulated waveguide, which are evaluated by employing a finite element discretization of Eqn.~\eqref{govEQ} and the application of Bloch conditions on a unit cell,~\cite{hussein2014dynamics} \textit{i.e.} $w(\omega,x+L_c)=w(\omega,x)e^{-i\kappa L_c}$, where $\kappa$ is the wavenumber. Figure~\ref{Fig2a} depicts two dispersion surfaces as a function of $\phi$ and $\mu=\kappa L_c$, which are separated by a gap of center frequency close to $9.7$ kHz. The inset displays the dispersion $\omega(\mu)$ for $\phi=0$, and shows five bands. The inset also highlights, through the shaded blue area, the frequency range corresponding to the surfaces shown in the main plot, which focuses primarily on the gap separating the fourth and fifth bands. The topology of the bands is described by the Chern number evaluated in the $[\mu,\phi]$ domain,~\cite{hatsugai1993chern,rosa2019edge,riva2020edge} which results in $C=1$ for the fourth band (green surface) of Fig.~\ref{Fig2a}. Similarly, a label for gap $r$ is assigned by the algebraic sum of the Chern number of the bands below it,~\cite{hatsugai1993chern,rosa2019edge,riva2020edge} \textit{i.e.} $C_g^{(r)}=\sum_{n=1}^rC_n$. The first gap (shaded gray region in Fig.~\ref{Fig2a}) is topologically trivial with $C_g=0$, as a result of the contribution of the three bands below, for which $C_1=1, C_2=-2, C_3=1$. The following gap, separating the two bands represented by the shaded red region in Fig.~\ref{Fig2a}, is non-trivial (see SM~\cite{SM} for further details on the dispersion and Chern number computation). The non-zero label $C_g=1$ of the gap in Fig.~\ref{Fig2a} indicates its ability to support an edge state spanning the gap in a finite structure.~\cite{rosa2019edge,riva2020edge} We illustrate this by computing the eigenfrequencies of a finite beam of length $L=57.6$ cm, comprising 8 unit cells for a total of $24$ pair of patches. Results for the beam in simply supported conditions are shown in Fig.~\ref{Fig2b}, which displays the variation of the eigenfrequencies as a function of $\phi$ (black lines), superimposed to the bulk bands represented by the shaded gray areas. The additional mode spanning the non-trivial gap is an edge state, where dashed (solid) lines are used for values of $\phi$ corresponding to left (right) localized modes. The three representative modes marked in Fig.~\ref{Fig2b} are displayed in Fig.~\ref{Fig2c} to illustrate a transition of the edge state from right-localized (I), to bulk (II), and then to left-localized (III) for increasing $\phi$ values. Such transition is hereafter employed to induce edge-to-edge pumping through a smooth modulation of $\phi$ in time.

\begin{figure*}[ht!]
	\centering
	\subfigure[]{\includegraphics[height=0.245\textwidth]{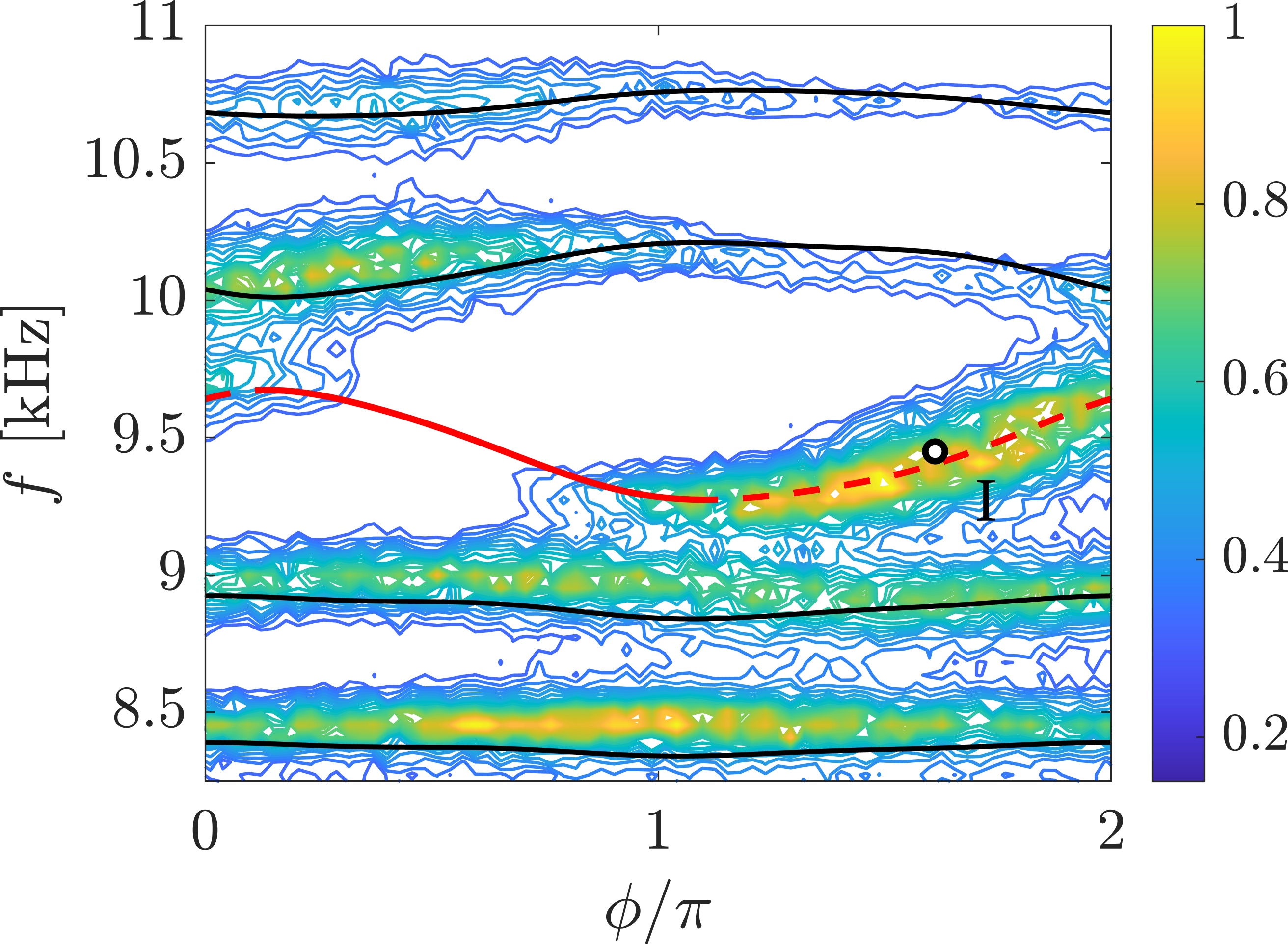}\label{Fig3a}}
	\subfigure[]{\includegraphics[height=0.245\textwidth]{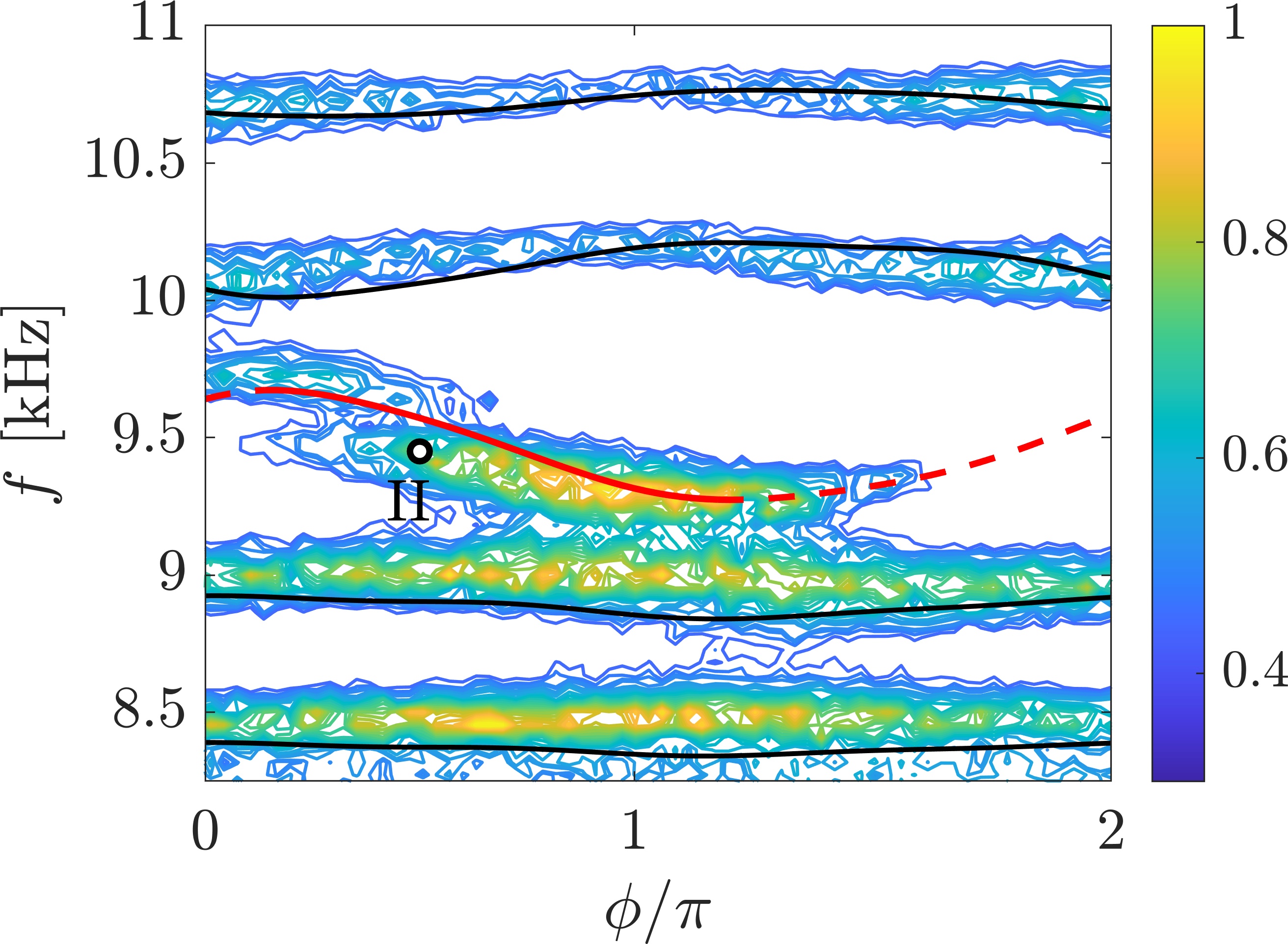}\label{Fig3b}}
	\subfigure[]{\includegraphics[height=0.245\textwidth]{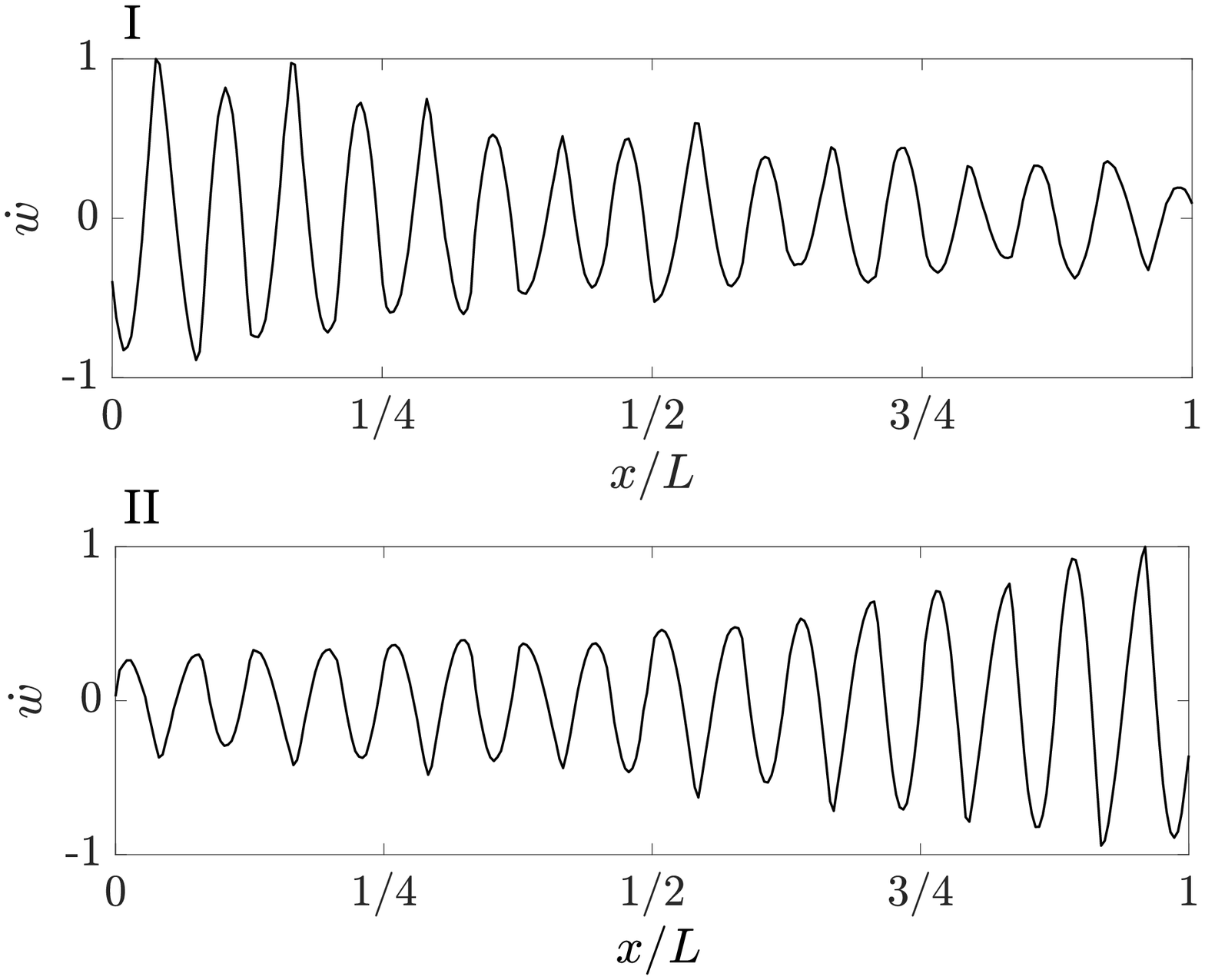}\label{Fig3c}}
	\caption{Experimental spectral characterization of modulated electromechanical beam. (a,b) Measured frequency response as a function of $\phi$ (contours) for excitation at the left (a) and right (b) boundary, superimposed to the eigenfrequencies of bulk (black) and edge (red) modes. The left excitation identifies mostly the left-localized portion of the branch of the edge state (dashed lines), while results the right excitation identifies the right-localized portion (solid line). (c) Representative experimental response for left (I) and right (II) localized modes, corresponding to points marked in (a) and (b), respectively.}
	\label{Fig3}
\end{figure*}

%\section{Experimental spectral characterization and temporal pumping}\label{experimentalsec}

\begin{figure*}[ht!]
	\centering
	\subfigure[]{\includegraphics[height=0.25\textwidth]{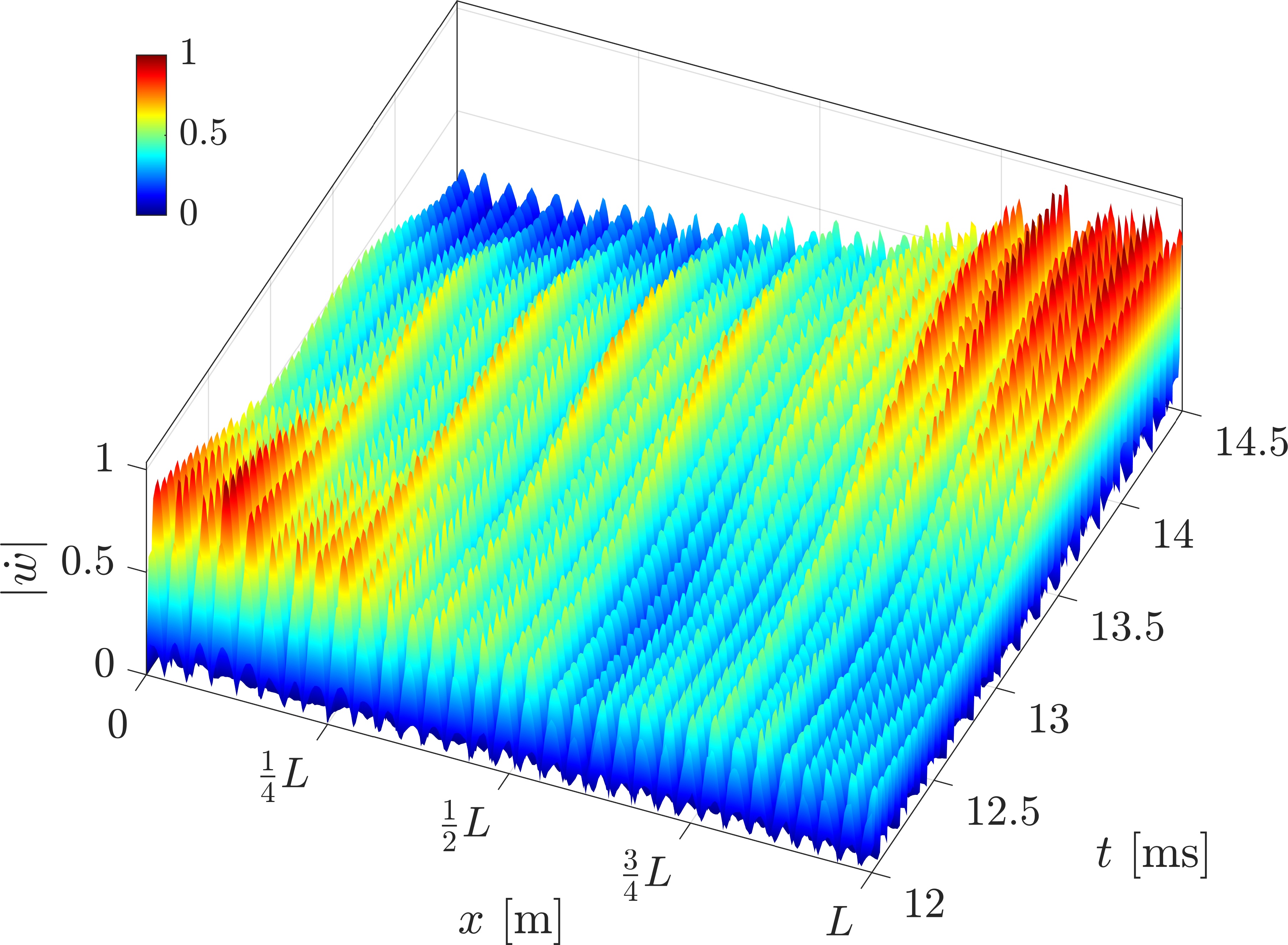}\label{Fig4a}}
	\subfigure[]{\includegraphics[height=0.25\textwidth]{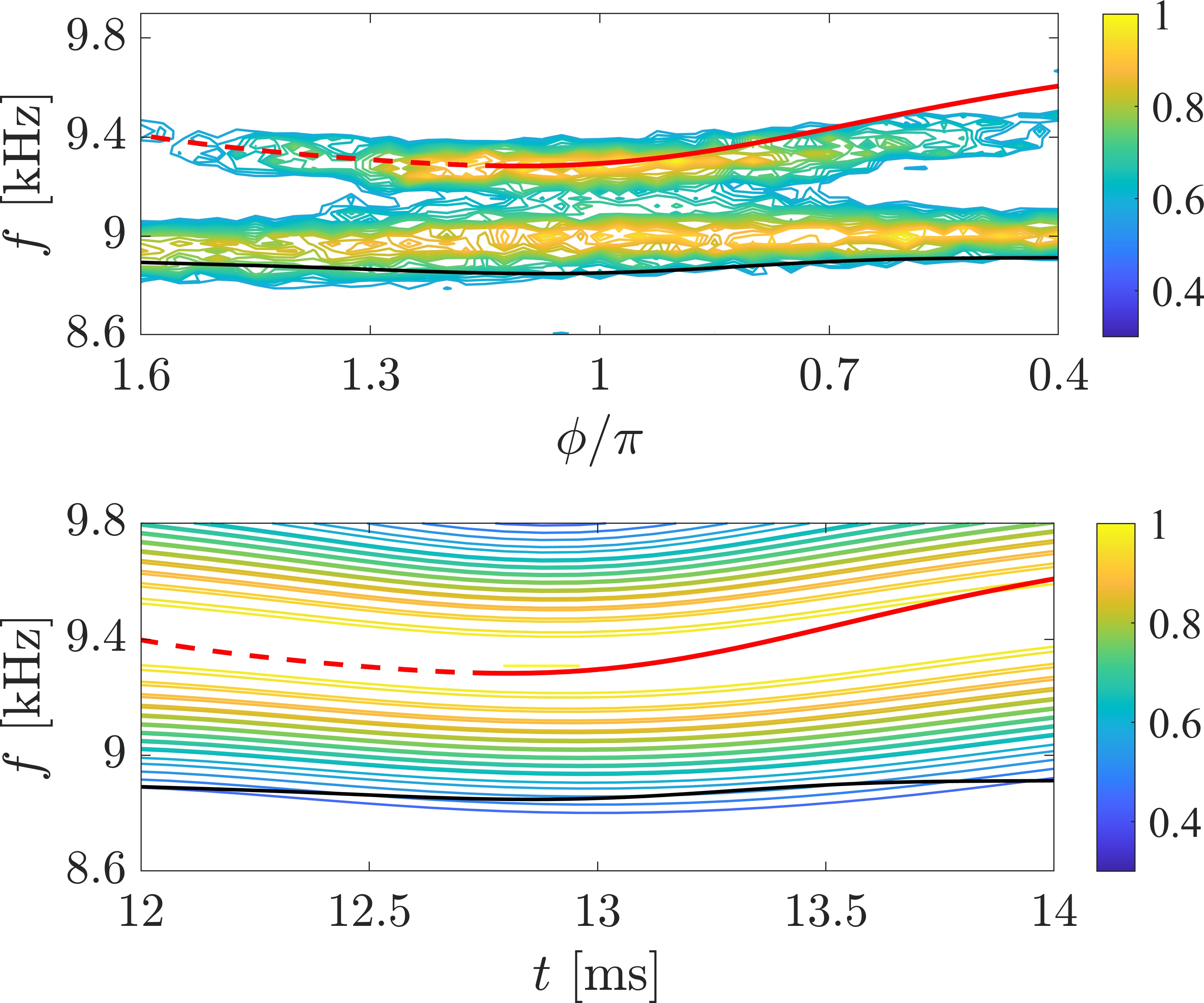}\label{Fig4b}}
	\subfigure[]{\includegraphics[height=0.25\textwidth]{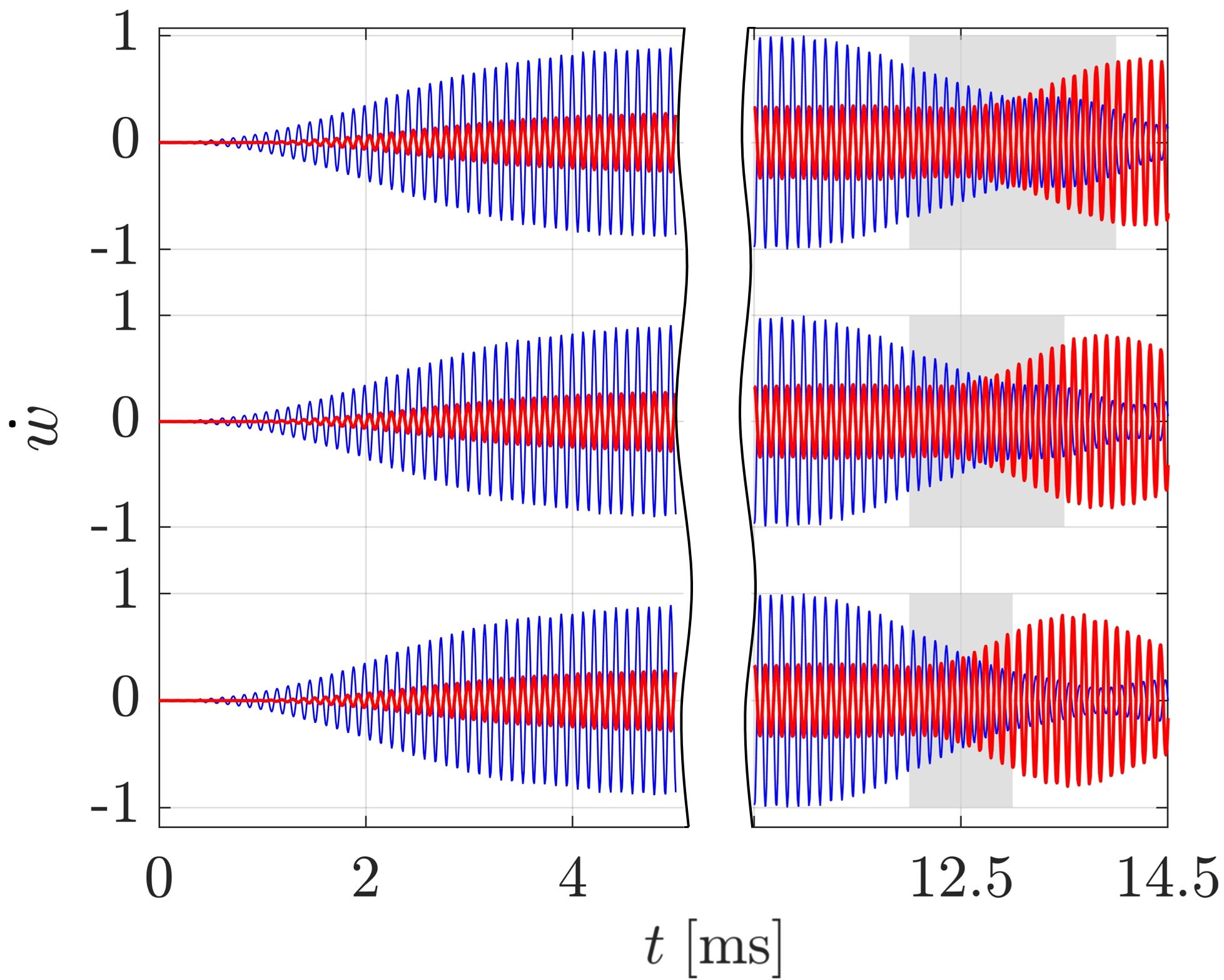}\label{Fig4c}}
	\caption{Experimental demonstration of temporal pumping in the electromechanical waveguide. (a) Transient time history illustrating a transition from a left-localized mode to a right-localized mode, induced by a linear temporal phase variation from $\phi_1=1.6\pi \to \phi_2=0.4\pi$ starting at $t=12$ ms, with a duration of $2$ ms. (b) Spectral content of broad-band excitation in quasi-static conditions (top) compared to spectrogram of the temporal pump (bottom), illustrating the adiabatic evolution along the branch of the edge state occurring in the pump with negligible influence of the neighboring bulk mode. (c) Signals at left (blue) and right (red) boundaries of the beam for temporal pumps induced within different modulation windows (shaded gray regions). In the initial $12$ ms, steady state vibrations of the left-localized mode are induced (with duration halved for better visualization), while different phase modulation durations (top: $2$ ms, middle: $1.5$ ms and bottom: $1$ ms) delay the arrival of the signal at the right end of the beam. }
	\label{Fig4}
\end{figure*}

The experimental investigations have as a first goal the characterization of the beam spectrum and its dependence on $\phi$. The waveguide, also comprising 24 pairs of piezoelectric patches, is clamped at both boundaries and excited at one end by one of the patches. A scanning laser Doppler vibrometer (LDV) measures the velocity field $\dot{w}(x,t)$ along the beam resulting from a band-limited noise excitation in the $3-15$ kHz frequency range. The signal is continuously applied for the duration of the test ($T=2.2$ s), while the phase $\phi$ varies in the interval $[0, 2\pi]$. The resulting input and output signals are post-processed to estimate the frequency response of the beam as a function of the phase $\phi$. To this end, the signals are multiplied by a rectangular window of length $T_s=0.22$ s, centered at an instant $t_0$, and the frequency response of the beam for $\phi=\phi(t_0)$, i.e. $W(x,\phi(t_0),\omega)$, is obtained by employing an H1 frequency estimator.~\cite{worden2019nonlinearity} The center of the window $t_0$ is smoothly translated in time, while the $L_2$ norm is taken along the spatial $x$ coordinate. This produces estimations of the frequency response as a function of $\phi(t_0)$, i.e $W(\phi,\omega)$ (additional details can be found in the SM~\cite{SM}). The results are reported as contour plots in Figs.~\ref{Fig3}(a,b), which correspond to two experiments where the beam is excited at the left and right boundary, respectively. Black and red lines superimposed to the experimental contours correspond to the eigenfrequencies of the bulk and edge modes predicted numerically. The experimental results show a good agreement with this numerical spectrum, which is obtained according to the procedure described in the SM.~\cite{SM} In the experiments, left excitation (Fig.~\ref{Fig3a}) reproduces mostly the left-localized branch of the edge state (dashed lines), while the right excitation experiment (Fig.~\ref{Fig3b}) captures primarily the right-localized branch (solid line). Experimentally measured left-localized and right-localized modes corresponding to the points marked as `I' and `II' Figs.~\ref{Fig3}(a,b) are shown in Figure~\ref{Fig3c}. 

Upon characterization of the spectrum and corresponding edge states, we next impose the smooth temporal variation of $\phi$ to induce topological pumping. We first target the left-localized mode defined for $\phi_1=1.6\pi$ (mode I in Figs.~\ref{Fig3}(a,c)) by applying a harmonic excitation of frequency $9.45$kHz to the left boundary. The excitation signal is maintained for an interval of $12$ ms, which is found sufficient to induce the steady-state motion of the left-localized mode. Upon stopping the excitation, we observe the free evolution of the waveguide response as the phase $\phi$ is varied linearly in time to reach a value of $\phi_2=0.4\pi$, which takes approximately $2$ ms (Fig.~\ref{Fig4b}). This modulation of the phase causes the transition to the right-localized mode (mode II in Figs.~\ref{Fig3}(b,c)). Figure~\ref{Fig4a} displays the magnitude of the experimentally recorded transverse motion of the beam from $t=12$ ms onwards, \emph{i.e.} after steady state conditions are reached. During the displayed time interval, the linear variation of the phase from $\phi_1=1.6\pi \to \phi_2= 0.4\pi$ induces the expected transition from a left-localized edge state to a right-localized state, as shown in Fig.~\ref{Fig4a}. As detailed in the SM,~\cite{SM} to better illustrate this transition, the recorded response is compensated for dissipation by extracting a temporal decay factor $\xi$ for the edge state at constant $\phi$, and then multiplying the transient time history $\dot{w}(x,t)$ by $e^{\xi t}$. The procedure does not alter the spatial distribution of the velocity field $\dot{w}$ at any given time instant, but allows for a better visualization of the pump and approximates the behavior of the system should dissipation be minimized. This can potentially be achieved in future studies by introducing a negative loss factor through suitable active circuits.~\cite{grinberg2020robust} The topological pump displayed in Fig.~\ref{Fig4a} is characterized by an adiabatic~\cite{nassar2018quantization} transition along the branch of the edge state, as illustrated by the spectral content in Fig.~\ref{Fig4b}. In the upper panel, the spectra in Figs.~\ref{Fig3}(a,b) are averaged to provide a single spectral characterization of the waveguide, and to highlight the presence of an edge state and of a bulk mode. The spectrogram in the bottom panel is obtained through the Fourier transformation and appropriate windowing of the transient pump displayed in Fig.~\ref{Fig4a} (see details in the SM~\cite{SM}). The results illustrate how energy remains concentrated around the edge state branch, with negligible contribution to the neighboring bulk modes (black line) as expected in an adiabatic state evolution.~\cite{nassar2018quantization}

Finally, we elucidate how the temporal pump realized with controllable phase modulation speeds can be of potential interest for the manipulation and transport of information across the waveguide. Figure~\ref{Fig4c} displays the velocity time history for a point at the left (blue) and right boundary (red) of the beam. The three plots correspond to edge-to-edge transitions driven by different modulation speeds. 
Under the aforementioned testing conditions, we employed the same input signal, whereby during the first $12$ ms a standing left-localized edge state is induced.
At $t=12$ ms the linear temporal phase modulation $\phi(t)$ starts, ranging from $\phi_1=1.6\pi$ to $\phi_2=0.4\pi$ during an interval of $1$ ms (bottom), $1.5$ ms (middle) and $2$ ms (top). The time duration of the phase modulation is highlighted by shaded gray areas to illustrate how the arrival time of the signal to the right end of the beam (in red) is controlled by the rate of phase modulation. 
This ability to control this arrival time independently from the underlying properties of the medium (the beam in this case) suggests opportunities for the designs of digitally controllable electromechanical delay lines based on topological pumping.

%\section{Conclusions}\label{Conclusionsec}
This letter illustrates the first experimental demonstration of temporal pumping in a continuous electromechanical waveguide with controllable modulation capabilities. Such modulations are employed for the topological pumping of edge states according to different modulation rates. This suggests the possibility to implement transfer of information in waveguides at speeds that are uniquely defined by the induced phase modulation. These results highlight potential applications to devices relying on robust signal transport, with tunable arrival times and phase delays, and open potential pathways for manipulating elastic waves using electromechanical waveguides. The setup is also a convenient platform for fundamental studies on higher order topologies using time as a synthetic dimension. 

\begin{acknowledgments}
The authors gratefully acknowledge the support from the National Science Foundation (NSF) through the EFRI 1741685 grant and from the Army Research office through grant W911NF-18-1-0036. The Italian Ministry of Education, University and Research is acknowledged for the support provided through the Project "Department of Excellence LIS4.0 - Lightweight and Smart Structures for Industry 4.0”.
\end{acknowledgments}

%\bibliographystyle{apsrev4-1}
\bibliographystyle{unsrt}
\bibliography{References}

\begin{thebibliography}{10}

\bibitem{hasan2010colloquium}
M~Zahid Hasan and Charles~L Kane.
\newblock Colloquium: topological insulators.
\newblock {\em Reviews of Modern Physics}, 82(4):3045, 2010.

\bibitem{lu2014topological}
Ling Lu, John~D Joannopoulos, and Marin Solja{\v{c}}i{\'c}.
\newblock Topological photonics.
\newblock {\em Nature Photonics}, 8(11):821, 2014.

\bibitem{khanikaev2013photonic}
Alexander~B Khanikaev, S~Hossein Mousavi, Wang-Kong Tse, Mehdi Kargarian,
  Allan~H MacDonald, and Gennady Shvets.
\newblock Photonic topological insulators.
\newblock {\em Nature materials}, 12(3):233, 2013.

\bibitem{PhysRevLett.114.114301}
Zhaoju Yang, Fei Gao, Xihang Shi, Xiao Lin, Zhen Gao, Yidong Chong, and Baile
  Zhang.
\newblock Topological acoustics.
\newblock {\em Phys. Rev. Lett.}, 114:114301, Mar 2015.

\bibitem{fleury2016floquet}
Romain Fleury, Alexander~B Khanikaev, and Andrea Alu.
\newblock Floquet topological insulators for sound.
\newblock {\em Nature communications}, 7:11744, 2016.

\bibitem{lu2017observation}
Jiuyang Lu, Chunyin Qiu, Liping Ye, Xiying Fan, Manzhu Ke, Fan Zhang, and
  Zhengyou Liu.
\newblock Observation of topological valley transport of sound in sonic
  crystals.
\newblock {\em Nature Physics}, 13(4):369, 2017.

\bibitem{mousavi2015topologically}
S~Hossein Mousavi, Alexander~B Khanikaev, and Zheng Wang.
\newblock Topologically protected elastic waves in phononic metamaterials.
\newblock {\em Nature communications}, 6:8682, 2015.

\bibitem{klitzing1980new}
K~v Klitzing, Gerhard Dorda, and Michael Pepper.
\newblock New method for high-accuracy determination of the fine-structure
  constant based on quantized hall resistance.
\newblock {\em Physical Review Letters}, 45(6):494, 1980.

\bibitem{thouless1982quantized}
David~J Thouless, Mahito Kohmoto, M~Peter Nightingale, and Md~den Nijs.
\newblock Quantized hall conductance in a two-dimensional periodic potential.
\newblock {\em Physical review letters}, 49(6):405, 1982.

\bibitem{prodan2009topological}
Emil Prodan and Camelia Prodan.
\newblock Topological phonon modes and their role in dynamic instability of
  microtubules.
\newblock {\em Physical review letters}, 103(24):248101, 2009.

\bibitem{wang2015topological}
Pai Wang, Ling Lu, and Katia Bertoldi.
\newblock Topological phononic crystals with one-way elastic edge waves.
\newblock {\em Physical review letters}, 115(10):104302, 2015.

\bibitem{nash2015topological}
Lisa~M Nash, Dustin Kleckner, Alismari Read, Vincenzo Vitelli, Ari~M Turner,
  and William~TM Irvine.
\newblock Topological mechanics of gyroscopic metamaterials.
\newblock {\em Proceedings of the National Academy of Sciences},
  112(47):14495--14500, 2015.

\bibitem{souslov2017topological}
Anton Souslov, Benjamin~C Van~Zuiden, Denis Bartolo, and Vincenzo Vitelli.
\newblock Topological sound in active-liquid metamaterials.
\newblock {\em Nature Physics}, 13(11):1091, 2017.

\bibitem{mitchell2018amorphous}
Noah~P Mitchell, Lisa~M Nash, Daniel Hexner, Ari~M Turner, and William~TM
  Irvine.
\newblock Amorphous topological insulators constructed from random point sets.
\newblock {\em Nature Physics}, 14(4):380, 2018.

\bibitem{chen2019mechanical}
H~Chen, LY~Yao, H~Nassar, and GL~Huang.
\newblock Mechanical quantum hall effect in time-modulated elastic materials.
\newblock {\em Physical Review Applied}, 11(4):044029, 2019.

\bibitem{susstrunk2015observation}
Roman S{\"u}sstrunk and Sebastian~D Huber.
\newblock Observation of phononic helical edge states in a mechanical
  topological insulator.
\newblock {\em Science}, 349(6243):47--50, 2015.

\bibitem{pal2016helical}
Raj~Kumar Pal, Marshall Schaeffer, and Massimo Ruzzene.
\newblock Helical edge states and topological phase transitions in phononic
  systems using bi-layered lattices.
\newblock {\em Journal of Applied Physics}, 119(8):084305, 2016.

\bibitem{PhysRevB.98.094302}
H.~Chen, H.~Nassar, A.~N. Norris, G.~K. Hu, and G.~L. Huang.
\newblock Elastic quantum spin hall effect in kagome lattices.
\newblock {\em Phys. Rev. B}, 98:094302, Sep 2018.

\bibitem{chaunsali2018subwavelength}
Rajesh Chaunsali, Chun-Wei Chen, and Jinkyu Yang.
\newblock Subwavelength and directional control of flexural waves in
  zone-folding induced topological plates.
\newblock {\em Physical Review B}, 97(5):054307, 2018.

\bibitem{PhysRevX.8.031074}
M.~Miniaci, R.~K. Pal, B.~Morvan, and M.~Ruzzene.
\newblock Experimental observation of topologically protected helical edge
  modes in patterned elastic plates.
\newblock {\em Phys. Rev. X}, 8:031074, Sep 2018.

\bibitem{pal2017edge}
Raj~Kumar Pal and Massimo Ruzzene.
\newblock Edge waves in plates with resonators: an elastic analogue of the
  quantum valley hall effect.
\newblock {\em New Journal of Physics}, 19(2):025001, 2017.

\bibitem{vila2017observation}
Javier Vila, Raj~Kumar Pal, and Massimo Ruzzene.
\newblock Observation of topological valley modes in an elastic hexagonal
  lattice.
\newblock {\em Physical Review B}, 96(13):134307, 2017.

\bibitem{liu2018tunable}
Ting-Wei Liu and Fabio Semperlotti.
\newblock Tunable acoustic valley--hall edge states in reconfigurable phononic
  elastic waveguides.
\newblock {\em Physical Review Applied}, 9(1):014001, 2018.

\bibitem{liu2019experimental}
Ting-Wei Liu and Fabio Semperlotti.
\newblock Experimental evidence of robust acoustic valley hall edge states in a
  nonresonant topological elastic waveguide.
\newblock {\em Physical Review Applied}, 11(1):014040, 2019.

\bibitem{qi2008topological}
Xiao-Liang Qi, Taylor~L Hughes, and Shou-Cheng Zhang.
\newblock Topological field theory of time-reversal invariant insulators.
\newblock {\em Physical Review B}, 78(19):195424, 2008.

\bibitem{kraus2016quasiperiodicity}
Yaacov~E Kraus and Oded Zilberberg.
\newblock Quasiperiodicity and topology transcend dimensions.
\newblock {\em Nature Physics}, 12(7):624, 2016.

\bibitem{ozawa2016synthetic}
Tomoki Ozawa, Hannah~M Price, Nathan Goldman, Oded Zilberberg, and Iacopo
  Carusotto.
\newblock Synthetic dimensions in integrated photonics: From optical isolation
  to four-dimensional quantum hall physics.
\newblock {\em Physical Review A}, 93(4):043827, 2016.

\bibitem{lee2018electromagnetic}
Ching~Hua Lee, Yuzhu Wang, Youjian Chen, and Xiao Zhang.
\newblock Electromagnetic response of quantum hall systems in dimensions five
  and six and beyond.
\newblock {\em Physical Review B}, 98(9):094434, 2018.

\bibitem{alvarez2019edge}
VM~Martinez Alvarez and MD~Coutinho-Filho.
\newblock Edge states in trimer lattices.
\newblock {\em Physical Review A}, 99(1):013833, 2019.

\bibitem{rosa2019edge}
Matheus~IN Rosa, Raj~Kumar Pal, Jos{\'e}~RF Arruda, and Massimo Ruzzene.
\newblock Edge states and topological pumping in spatially modulated elastic
  lattices.
\newblock {\em Physical Review Letters}, 123(3):034301, 2019.

\bibitem{kraus2012topological}
Yaacov~E Kraus, Yoav Lahini, Zohar Ringel, Mor Verbin, and Oded Zilberberg.
\newblock Topological states and adiabatic pumping in quasicrystals.
\newblock {\em Physical review letters}, 109(10):106402, 2012.

\bibitem{apigo2019observation}
David~J Apigo, Wenting Cheng, Kyle~F Dobiszewski, Emil Prodan, and Camelia
  Prodan.
\newblock Observation of topological edge modes in a quasiperiodic acoustic
  waveguide.
\newblock {\em Physical review letters}, 122(9):095501, 2019.

\bibitem{ni2019observation}
Xiang Ni, Kai Chen, Matthew Weiner, David~J Apigo, Camelia Prodan, Andrea
  Al{\`u}, Emil Prodan, and Alexander~B Khanikaev.
\newblock Observation of hofstadter butterfly and topological edge states in
  reconfigurable quasi-periodic acoustic crystals.
\newblock {\em Communications Physics}, 2(1):55, 2019.

\bibitem{Pal_2019}
Raj~Kumar Pal, Matheus I~N Rosa, and Massimo Ruzzene.
\newblock Topological bands and localized vibration modes in quasiperiodic
  beams.
\newblock {\em New Journal of Physics}, 21(9):093017, sep 2019.

\bibitem{xia2020topological}
Yiwei Xia, Alper Erturk, and Massimo Ruzzene.
\newblock Topological edge states in quasiperiodic locally resonant
  metastructures.
\newblock {\em Physical Review Applied}, 13(1):014023, 2020.

\bibitem{zilberberg2018photonic}
Oded Zilberberg, Sheng Huang, Jonathan Guglielmon, Mohan Wang, Kevin~P Chen,
  Yaacov~E Kraus, and Mikael~C Rechtsman.
\newblock Photonic topological boundary pumping as a probe of 4d quantum hall
  physics.
\newblock {\em Nature}, 553(7686):59, 2018.

\bibitem{lohse2018exploring}
Michael Lohse, Christian Schweizer, Hannah~M Price, Oded Zilberberg, and
  Immanuel Bloch.
\newblock Exploring 4d quantum hall physics with a 2d topological charge pump.
\newblock {\em Nature}, 553(7686):55, 2018.

\bibitem{petrides2018six}
Ioannis Petrides, Hannah~M Price, and Oded Zilberberg.
\newblock Six-dimensional quantum hall effect and three-dimensional topological
  pumps.
\newblock {\em Physical Review B}, 98(12):125431, 2018.

\bibitem{verbin2015topological}
Mor Verbin, Oded Zilberberg, Yoav Lahini, Yaacov~E Kraus, and Yaron Silberberg.
\newblock Topological pumping over a photonic fibonacci quasicrystal.
\newblock {\em Physical Review B}, 91(6):064201, 2015.

\bibitem{riva2020edge}
Emanuele Riva, Matheus~IN Rosa, and Massimo Ruzzene.
\newblock Edge states and topological pumping in stiffness-modulated elastic
  plates.
\newblock {\em Physical Review B}, 101(9):094307, 2020.

\bibitem{nakajima2016topological}
Shuta Nakajima, Takafumi Tomita, Shintaro Taie, Tomohiro Ichinose, Hideki
  Ozawa, Lei Wang, Matthias Troyer, and Yoshiro Takahashi.
\newblock Topological thouless pumping of ultracold fermions.
\newblock {\em Nature Physics}, 12(4):296, 2016.

\bibitem{lohse2016thouless}
Michael Lohse, Christian Schweizer, Oded Zilberberg, Monika Aidelsburger, and
  Immanuel Bloch.
\newblock A thouless quantum pump with ultracold bosonic atoms in an optical
  superlattice.
\newblock {\em Nature Physics}, 12(4):350, 2016.

\bibitem{grinberg2020robust}
Inbar~Hotzen Grinberg, Mao Lin, Cameron Harris, Wladimir~A Benalcazar,
  Christopher~W Peterson, Taylor~L Hughes, and Gaurav Bahl.
\newblock Robust temporal pumping in a magneto-mechanical topological
  insulator.
\newblock {\em Nature communications}, 11(1):1--9, 2020.

\bibitem{brouzos2019non}
I~Brouzos, I~Kiorpelidis, FK~Diakonos, and G~Theocharis.
\newblock Non-adiabatic time-optimal edge mode transfer on mechanical
  topological chain.
\newblock {\em arXiv preprint arXiv:1911.03375}, 2019.

\bibitem{longhi2019topological}
Stefano Longhi.
\newblock Topological pumping of edge states via adiabatic passage.
\newblock {\em Physical Review B}, 99(15):155150, 2019.

\bibitem{riva2020adiabatic}
Emanuele Riva, Vito Casieri, Ferruccio Resta, and Francesco Braghin.
\newblock Adiabatic pumping via avoided crossings in stiffness modulated
  quasiperiodic beams.
\newblock {\em arXiv preprint arXiv:2003.11525}, 2020.

\bibitem{nassar2018quantization}
H~Nassar, H~Chen, AN~Norris, and GL~Huang.
\newblock Quantization of band tilting in modulated phononic crystals.
\newblock {\em Physical Review B}, 97(1):014305, 2018.

\bibitem{trainiti2019time}
Giuseppe Trainiti, Yiwei Xia, Jacopo Marconi, Gabriele Cazzulani, Alper Erturk,
  and Massimo Ruzzene.
\newblock Time-periodic stiffness modulation in elastic metamaterials for
  selective wave filtering: theory and experiment.
\newblock {\em Physical review letters}, 122(12):124301, 2019.

\bibitem{marconi2020experimental}
Jacopo Marconi, Emanuele Riva, Matteo Di~Ronco, Gabriele Cazzulani, Francesco
  Braghin, and Massimo Ruzzene.
\newblock Experimental observation of nonreciprocal band gaps in a
  space-time-modulated beam using a shunted piezoelectric array.
\newblock {\em Physical Review Applied}, 13(3):031001, 2020.

\bibitem{graff2012wave}
Karl~F Graff.
\newblock {\em Wave motion in elastic solids}.
\newblock Courier Corporation, 2012.

\bibitem{SM}
See supplemental material at xxxx for more details on the simulation
  procedures, experimental setup and methodology.

\bibitem{hussein2014dynamics}
Mahmoud~I Hussein, Michael~J Leamy, and Massimo Ruzzene.
\newblock Dynamics of phononic materials and structures: Historical origins,
  recent progress, and future outlook.
\newblock {\em Applied Mechanics Reviews}, 66(4), 2014.

\bibitem{hatsugai1993chern}
Yasuhiro Hatsugai.
\newblock Chern number and edge states in the integer quantum hall effect.
\newblock {\em Physical review letters}, 71(22):3697, 1993.

\bibitem{worden2019nonlinearity}
Keith Worden.
\newblock {\em Nonlinearity in structural dynamics: detection, identification
  and modelling}.
\newblock CRC Press, 2019.

\end{thebibliography}
\end{document}